# Poisson-Lie T-Duality in Supersymmetric WZNW Model

F. Assaoui, N. Benhamou and T. Lhallabi*

*Section of High Energy Physics, H. E. P. L, University Mohammed V,*

*Scientific Faculty, Rabat, Morocco\*\**

*and*

*The Abdus Salam International Centre for Theoretical Physics, Trieste, Italy.*

## Abstract

The description of the two sets of (4,0) supersymmetric models that are related by non-abelian duality transformations is given. The (4,0) supersymmetric WZNW is constructed and the formulation of the (4,0) supersymmetric sigma model dual to (4,0) supersymmetric WZNW model in the sense of Poisson-Lie T-duality is described.

*

\* Regular Associate of the abdus Salam ICTP.
\*\* Perment address. E-mail: Lhallabi@fsr.ac.ma

## 1 - Introduction

The WZNW models are the most important models of two-dimensional field theory due to their applications in string theory and to the fact that many other interesting models can be understood as their reductions [1]. Furthermore, some conformal field theory results have been obtained for open strings in SU(2) WZNW model in [2] and the arbitrariness in defining the WZNW for open strings is completely fixed by the requirement of the Poisson-Lie symmetry [3]. Moreover, for various Drinfeld doubles underlying the structure of the Poisson-Lie T-duality, the same WZNW model is recovered [4,3]. On the other hand, it was shown in [5,3] that WZNW models on the compact groups are the natural examples of Poisson-Lie dualizable sigma models. The supersymmetric generalization of Poisson-Lie T-duality was considered in [6] and the dual pairs of N = 2 superconformal WZNW models are constructed [7,4].

In the first part, of this paper, we consider a class of deformed (4,0) supersymmetric principal models where its non-abelian dual connects the SU(2) principal (4,0) supersymmetric model and the O(3) (4,0) supersymmetric sigma model. In the second part, we construct the (4,0) supersymmetric WZNW model and we show that the Poisson-Lie dualizable (4,0) supersymmetric sigma model is the standard (4,0) supersymmetric WZNW model. The detailed classical account of Poisson-Lie T-duality is given.

The paper is organized as follows: In section 2, we describe in some details the two sets of (4,0) supersymmetric models which are related to each other by non-abelian duality transformations. In section 3, we construct the (4,0) supersymmetric WZNW model and we describe the formulation of the (4,0) supersymmetric sigma model to (4,0) supersymmetric WZNW model in the sense of Poisson-Lie T-duality. Finally, in section 4, we make concluding remarks and discuss our results.

## 2 - (4,0) Supersymmetric Dually Related Models

Let us consider a superworld-sheet parameterized by the light cone coordinates $x^{\pm\pm}$, Grassman coordinates $\theta^{+}_{-\alpha}$ and harmonic variables $U^{\pm}_{\beta}$, which constitute the (4,0) analytic subspace [8]. The spinor and harmonic covariant derivatives on the (4,0) analytic subspace are given by

$$D^-_{+\alpha} = \frac{\partial}{\partial \theta^{+\alpha}_-} - 2i\theta^-_{-\alpha}\partial_{++}$$

$$D^-_{+\alpha} = \frac{\partial}{\partial \theta^{-\alpha}_-} \qquad (2.1)$$

$$D^{++} = d^{++} - 2\theta^+_-\theta^+_-\partial_{++}$$

where

$$d^{++} = U^{+\beta}\frac{\partial}{\partial U^{-\beta}}$$

We choose the original (4,0) supersymmetric model from the class of deformed (4,0) supersymmetric principal models, where the lagrangian can be written as

$$L = \frac{1}{2}R_{ab}L^a_i L^b_j D^{++}\Omega^i \partial_{--}\Omega^j \qquad (2.2)$$

where $R_{ab}$ is a constant, $\Omega^i$, i = 1, ..., dimG, which parametrizes the element G of a group G is an analytic superfield satisfying

$$D^+_+\Omega^i = 0 \qquad (2.3)$$

and $L^a_i$ denote the components of the left invariant Cartan-Maurer one form namely

$$L^a_i = \frac{1}{w}\text{tr}\left(\lambda^a G^{-1}\frac{\partial G}{\partial \Omega^i}\right) \qquad (2.4)$$

$\lambda^a$ are the generators of the Lie algebra of G such that

$$[\lambda^a, \lambda^b] = f^{abc}\lambda^c, \qquad (2.5)$$

and normalized according to

$$\text{tr}(\lambda^a\lambda^b) = w\,\delta^{ab} \qquad (2.6)$$

The lagrangian (2.2) can be rewritten as

$$L = \frac{1}{2}R_{ab}L^{++a}L^b_{--} \qquad (2.7)$$

where

$$L^{++a} = L_i^a D^{++}\Omega^i$$
$$L_{--}^a = L_i^a \partial_{--}\Omega^i \quad (2.8)$$

By using the equation

$$\partial_k L_i^a - \partial_i L_k^a + f^{acd} L_k^c L_i^d = 0, \quad (2.9)$$

which is deduced from the definition (2.4) and which is equivalent to the zero curvature equations

$$D^{++}L_{--}^a - \partial_{--}L^{++a} + f^{abc}L^{++b}L_{--}^c = 0, \quad (2.10)$$

we obtain the following equation of motion

$$R_{ab}(D^{++}L_{--}^a + \partial_{--}L^{++a}) - R_{ab}f^{acd}(L^{++c}L_{--}^b + L^{++b}L_{--}^c) = 0 \quad (2.11)$$

Thereafter, the lagrangian $L$ is invariant under the left G transformations, namely

$$\delta G = hG \quad (2.12)$$

where h is an element of G. Consequently, the conserved Noether supercurrents are given by

$$J^{++a} = R_{cd} N_a^c L^{++d}$$
$$J_{--}^a = R_{cd} N_a^c L_{--}^d \quad (2.13)$$

with

$$N_a^c = \frac{1}{w} tr(\lambda^c G^{-1} \lambda_a G) \quad (2.14)$$

The conservation equation is as follows

$$D^{++}J_{--}^a + \partial_{--}J^{++a} = 0 \quad (2.15)$$

On the other hand, in order to obtain the non-abelian dual of the model (2.2) we gauge the symmetry (2.12) and add a lagrangian multiplier term as in ref [9].

$$L_g = L + \frac{1}{2}(J^{++a}V_{--a} + J_{--}^a V_a^{++} + R_{ab}V^{++a}V_{--}^b) + \frac{1}{2}\chi^a(\partial_{--}V_a^{++} - D^{++}V_{--a} + f_a^{bc}V_b^{++}V_{--c})$$
$$(2.16)$$

The elimination of the lagrangian multiplier parameter $\chi^a$ leads to the following constraint

$$\partial_{--}V_a^{++} - D^{++}V_{--a} + f_a^{bc}V_b^{++}V_{--c} = 0 \qquad (2.17)$$

Furthermore, the equations of motion of the gauge superfields $V^{++}$ and $V_{--}$ allow to obtain

$$\begin{aligned}
V_{--}^a &= (M^{-1})^{ac}[\partial_{--}\chi^c - J_{--}^C] \\
V^{++a} &= -(M^{-1})^{ca}[D^{++}\chi^c + J^{++c}]
\end{aligned} \qquad (2.18)$$

with

$$M^{bc} = R^{bc} + f^{bca}\chi_a \qquad (2.19)$$

$$(M^{-1})^{ca}M^{bc} = \delta^{ac}$$

The use of the equations (2.18) in $L_g$ leads to the dual lagrangian of (2.2) namely

$$\tilde{L} = \frac{1}{2}D^{++}\chi_a(M^{-1})^{ab}\partial_{--}\chi_b \qquad (2.20)$$

where the gauge choice $G = 1$ is taken into account. The transformation connecting $L$ and $\tilde{L}$ is a particular case of the Poisson-Lie T-duality [10,11] when the dual group is abelian.

However, we note that the deformed (4,0) supersymmetric principal model (2.2) with $G = SU(2)$ can be reduced to $O(3)$ (4,0) supersymmetric sigma model. Indeed, by choosing $G = SU(2)$ and

$$R_{ab} = \frac{1}{\lambda}\begin{pmatrix} 1 & 0 & 0 \\ 0 & 1 & 0 \\ 0 & 0 & 1+g \end{pmatrix} \qquad (2.21)$$

where $\lambda$ is a coupling constant, g is the deformation parameter and by parametrizing an element G of SU(2) by

$$G = e^{\phi\tau^3} e^{\eta\tau^1} e^{\psi\tau^3} \qquad (2.22)$$

where $(\phi,\psi,\eta)$ are the Euler angles, $\tau$ the Pauli matrices, the expressions (2.8) become

$$L^{++1} = (D^{++}\eta)\cos\psi + (D^{++}\phi)\sin\eta \sin\psi$$
$$L^{++2} = -(D^{++}\eta)\sin\psi + (D^{++}\phi)\sin\eta \cos\psi$$
$$L^{++3} = (D^{++}\psi) + (D^{++}\phi)\cos\eta$$
$$L^{1}_{--} = (\partial_{--}\eta)\cos\psi + (\partial_{--}\phi)\sin\eta \sin\psi \qquad (2.23)$$
$$L^{2}_{--} = -(\partial_{--}\eta)\sin\psi + (\partial_{--}\phi)\sin\eta \cos\psi$$
$$L^{3}_{--} = (\partial_{--}\psi) + (\partial_{--}\phi)\cos\eta$$

Then the lagrangian of the deformed (4,0) supersymmetric SU(2) model (2.2) is explicitly given by

$$L = \frac{1}{2\lambda}\{(D^{++}\eta)\partial_{--}\eta + (1+g\cos^2\eta)D^{++}\phi\partial_{--}\phi - (1+g)D^{++}\psi \partial_{--}\psi \qquad (2.24)$$
$$+ (1+g)(D^{++}\psi (\partial_{--}\phi) + D^{++}\phi \partial_{--}\psi)\cos\eta\}$$

We remark that for g = -1, the $\psi$ superfield decouples and the expression (2.24) reduces to the lagrangian of the O(3) supersymmetric sigma model. As for the ordinary case [9], the lagrangian (2.24) describes a one parameter family of models relating SU(2) principal model (g = 0) and the O(3) one.

Furthermore, since $M_{ab}$ is not singular for g = -1, the dual (4,0) supersymmetric model is defined for all $g \geq -1$. In fact, rescaling the $\chi^a$ superfield as

$$\chi^a \to \frac{\chi^a}{\lambda} \qquad (2.25)$$

we obtain

$$\tilde{L} = \frac{1}{2\lambda^2} D^{++}\chi_a (M^{-1})^{ab} \partial_{--}\chi^b \qquad (2.26)$$

with

$$M^{ab} = \frac{1}{\lambda}\begin{pmatrix} 1 & \chi^3 & -\chi^2 \\ -\chi^3 & 1 & \chi^1 \\ \chi^2 & -\chi^1 & 1+g \end{pmatrix} \qquad (2.27)$$

So we have

$$(M^{-1}) = \frac{1}{\det M} M^{ct} \qquad (2.28)$$

where $M^{ct}$ means the transposition co-matrix which is given by

$$M^c = \frac{1}{\lambda^2} \begin{pmatrix} 1+g+\chi_1^2 & (1+g)\chi_3+\chi_1\chi_2 & \chi_1\chi_3-\chi_2 \\ -(1+g)\chi_3+\chi_1\chi_2 & (1+g)\chi_2^2 & \chi_1+\chi_2\chi_3 \\ \chi_2+\chi_1\chi_3 & -\chi_1+\chi_2\chi_3 & 1+\chi_3^2 \end{pmatrix} \quad (2.29)$$

and

$$\det M = \frac{1}{\lambda^3}\left[(1+g)(1+\chi_3^2)+\chi_1^2+\chi_2^2\right] = \frac{D}{\lambda^3}$$

Now introducing the variables

$$\chi_1 = \rho\cos\alpha, \qquad \chi_2 = \rho\sin\alpha, \qquad \chi_3 = z \quad (2.30)$$

we find that

$$M^{-1} = \frac{\lambda}{D}\begin{pmatrix} 1+g+\rho^2\cos^2\alpha & -(1+g)z+\rho^2\cos\alpha\sin\alpha & \rho\sin\alpha+\rho z\cos\alpha \\ (1+g)z+\rho^2\cos\alpha\sin\alpha & 1+g+\rho^2\sin^2\alpha & \rho z\sin\alpha-\rho\cos\alpha \\ \rho z\cos\alpha-\rho\sin\alpha & \rho\cos\alpha+\rho z\sin\alpha & 1+z^2 \end{pmatrix} \quad (2.31)$$

with

$$D = (1+g)(1+z^2)+\rho^2$$

Consequently, the dual lagrangian is given by

$$\tilde{L} = \frac{1}{2\lambda}\{(1+g+\rho^2)D^{++}\rho\,\partial_{--}\rho+(1+g)\rho^2 D^{++}\alpha\,\partial_{--}\alpha+$$
$$(1+z^2)D^{++}z\,\partial_{--}z+(1+g)z\rho(D^{++}\alpha\,\partial_{--}\rho-D^{++}\rho\,\partial_{--}\alpha)+ \quad (2.32)$$
$$\rho z(D^{++}\rho\,\partial_{--}z+D^{++}z\,\partial_{--}\rho)+\rho^2(D^{++}z\,\partial_{--}\alpha-D^{++}\alpha\,\partial_{--}\rho)\}$$

Note that for g = -1, the $\alpha$ and $\rho$ superfields are decoupled and the dual lagrangian (2.32) becomes

$$\tilde{L} = \frac{1}{2\lambda}\{D^{++}\rho\,\partial_{--}\rho+\frac{1+z^2}{\rho^2}D^{++}z\,\partial_{--}z+\frac{z}{\rho}(D^{++}\rho\,\partial_{--}z+D^{++}z\,\partial_{--}\rho)\} \quad (2.33)$$

which is the non-abelian dual of the O(3) (4,0) supersymmetric sigma model. This model can be also obtained by considering the (4,0) supersymmetric WZNW models, which we construct in the following section we construct and describe their Poisson-Lie T-duality.

## 3 - (4,0) SWZNW Models

We construct a (4,0) supersymmetric WZNW model by parametrizing the superworldsheet as the previous section and by using the analytic superfield G which takes values in a Lie group G. We assume that the Lie superalgebras g of the group G is equipped with ad-invariant non degenerate inner product $\langle \, , \, \rangle$. Furthermore, the analytic superfield G is expanded in terms of $\theta_-^+$ components as

$$G = g + \theta_-^+ \psi_+^- + \theta_-^+ \theta_-^+ F_{++}^{--} \tag{3.1}$$

where its inverse is defined from the relation

$$G^{-1} G = 1 \tag{3.2}$$

and is given by

$$G^{-1} = g^{-1} + \theta_-^+ \left[ -g^{-1} \psi_+^- g^{-1} \right] - \theta_-^+ \theta_-^+ \, g^{-1} \left[ F_{++}^{--} + \frac{1}{2} \psi_+^- w^{-1} \psi_+^- \right] g^{-1} \tag{3.3}$$

The action of the (4,0) supersymmetric WZNW model is explicitly written as

$$S_{WZNW}^{(4,0)} = \int d\mu_{--}^{++} \left\{ \langle H^{++}, H_{--} \rangle \right\} - \int d\mu_{--}^{++} dt \left\{ \left\langle G^{-1} \frac{\partial G}{\partial t} \, , \, [H^{++}, H_{--}] \right\rangle \right\} \tag{3.4}$$

where $d\mu_{--}^{++} = d^2 x \, d^2 \theta_-^+ dU$ is the measure of the (4,0) analytic subspace and

$$\begin{aligned} H^{++} &= G^{-1} D^{++} G \\ H_{--} &= G^{-1} \partial_{--} G \end{aligned} \tag{3.5}$$

As for the N = 1 supersymmetric WZNW case [4], the classical equations of motion can be obtained by making the variation of the action (3.4) which leads to

$$\partial_{--} H^{++} + D^{++} H_{--} = [H^{++}, H_{--}] \tag{3.6}$$

Thereafter, by taking into account the kinematics relation, which is obtained from (3.5), namely

$$\partial_{--} H^{++} - D^{++} H_{--} = [H^{++}, H_{--}] \tag{3.7}$$

we obtain the following constraints

$$\partial_{--} H^{++} = [H^{++}, H_{--}]$$
$$D^{++} H_{--} = 0 \quad (3.8)$$

By using these relations we obtain the following (4,0) supersymmetric Polyakov-Weigman formula [12]

$$S_{WZNW}^{(4,0)}[G_1 G_2] = S_{WZNW}^{(4,0)}[G_1] + S_{WZNW}^{(4,0)}[G_2] + \int d\mu_{++}^{--} \{\langle H_1^{++}, H_{2--}\rangle + \langle H_2^{++}, H_{1--}\rangle\} \quad (3.9)$$

However, for the description of the Poisson-Lie T-duality in (4,0) supersymmetric WZNW model we consider, as for the N=2 supersymmetric case [4], the correspondence between complex Manin triple (**g**$^c$, **g**$_1$, **g**$_2$) which is equipped with a hermitian conjugation $\tau$ [13] such that

$$\tau : G_{1,2} \to G_{2,1}$$

and the real Lie algebra **g** with non degenerate invariant inner product and skew-symmetric complex structures [4] for which the complexification **g**$^c$ of **g** is associated [4]. The Lie group version of this triple is the double Lie group (**G**$^c$, **G**$_1$, **G**$_2$) [14,15] and the real Lie group G is extracted from its complexification by the use of the hermitian conjugation $\tau$

$$G = \{G \in G^c / \tau(G) = G^{-1}\}$$

Furthermore, each element G of G$^c$ from the vicinity of its unit element admits the following decompositions

$$G = G_1 G_2^{-1} = \tilde{G}_2 \tilde{G}_1^{-1} \quad (3.10)$$

where $\tilde{G}_{1,2}$ are dressed transformed elements of $G_{1,2}$ [15]. Therefore, by using the decomposition (3.10) and the Polyakov-Weigman formula (3.9), the (4,0) supersymmetric WZNW action (3.4) becomes

$$S_{WZNW}^{(4,0)} = -\frac{1}{2} \int d\mu_{++}^{--} \{\langle \Gamma_1^{++}, \Gamma_{2--}\rangle + \langle \Gamma_2^{++}, \Gamma_{1--}\rangle + \langle \tilde{\Gamma}_2^{++}, \tilde{\Gamma}_{1--}\rangle + \langle \tilde{\Gamma}_1^{++}, \tilde{\Gamma}_{2--}\rangle\} \quad (3.11)$$

with

$$\Gamma^{++}_{1,2} = G^{-1}_{1,2} D^{++} G_{1,2} \quad , \quad \Gamma_{1,2--} = G^{-1}_{1,2} \partial_{--} G_{1,2}$$
$$\tilde{\Gamma}^{++}_{1,2} = \tilde{G}^{-1}_{1,2} D^{++} \tilde{G}_{1,2} \quad , \quad \tilde{\Gamma}_{1,2--} = \tilde{G}^{-1}_{1,2} \partial_{--} \tilde{G}_{1,2}$$
(3.12)

which correspond to the left invariant 1-forms on $G_{1,2}$, namely

$$\Gamma_{1,2} = G^{-1}_{1,2} dG_{1,2}$$
$$\tilde{\Gamma}_{1,2} = \tilde{G}^{-1}_{1,2} d\tilde{G}_{1,2}$$
(3.13)

On the other hand, we identify the Lie superalgebra $\mathbf{g^c}$ with the space of complex left invariant vector superfields of the group G. Thereafter, the basis in the Lie subsuperalgebras $\mathbf{g_1}$ and $\mathbf{g_2}$ are respectively given by $e_i$ and $e^i$, $i = 1, \ldots, d$ such that

$$\langle e^i , e_j \rangle = \delta^i_j$$
(3.14)

Then, we have

$$\Gamma_1 = \Gamma^i e_i \quad , \quad \tilde{\Gamma}_1 = \tilde{\Gamma}^i e_i$$
$$\Gamma_2 = \Gamma_i e^i \quad , \quad \tilde{\Gamma}_2 = \tilde{\Gamma}_i e^i$$
(3.15)

where the components of $\mathbf{g_2}$ valued 1-forms $\Gamma_2, \tilde{\Gamma}_2$ can be rewritten on the basis $e_i$ and $e^i$ as follows

$$\Gamma_i = E_{i\tilde{j}} \tilde{\Gamma}^j + E_{ij} \Gamma^j$$
$$\tilde{\Gamma}_i = E_{j\tilde{i}} \Gamma^j + E_{\tilde{i}\tilde{j}} \tilde{\Gamma}^j$$
(3.16)

Consequently, the (4,0) supersymmetric WZNW takes the following form

$$S^{(4,0)}_{WZNW} = -\frac{1}{2} \int d\mu^{--}_{++} \left\{ \Gamma^{++i} \Gamma_{--i} + \Gamma^{++}_i \Gamma^i_{--} + \tilde{\Gamma}^{++}_i \tilde{\Gamma}^i_{--} + \tilde{\Gamma}^{++i} \tilde{\Gamma}_{--i} \right\}$$
(3.17)

which, after the use of (3.16), becomes

$$S^{(4,0)}_{WZNW} = -\frac{1}{2} \int d\mu^{--}_{++} \left\{ K_{ij} \Gamma^{++i} \Gamma^j_{--} + K_{i\tilde{j}} (\Gamma^{++i} \tilde{\Gamma}^j_{--} + \tilde{\Gamma}^{++j} \Gamma^i_{--}) + K_{\tilde{i}\tilde{j}} \tilde{\Gamma}^{++i} \tilde{\Gamma}^j_{--} \right\}$$
(3.18)

where
$$K_{ij} = E_{ij} + E_{ji}$$
$$K_{i\tilde{j}} = 2E_{i\tilde{j}} \quad (3.19)$$
$$K_{\tilde{i}\tilde{j}} = E_{\tilde{i}\tilde{j}} + E_{\tilde{j}\tilde{i}}$$

From these formulas, we can consider the (4,0) supersymmetric WZNW model as a (4,0) supersymmetric sigma model on the complex Lie group $G_1$. Furthermore, in order to obtain the Poisson-Lie symmetry conditions, let us derive the equations of motion by making a variation of the following left group action

$$\delta G_1 = \varepsilon_1^i \vartheta_i$$
$$\delta \tilde{G}_1 = \tilde{\varepsilon}_1^i \tilde{\vartheta}_i \quad (3.20)$$

These are given by

$$D^{++}V_{--i} + \partial_{--}V_i^{++} - (L_{\vartheta_i}K)_{--}^{++} = 0$$
$$D^{++}\tilde{V}_{--i} + \partial_{--}\tilde{V}_i^{++} - (L_{\tilde{\vartheta}_i}K)_{--}^{++} = 0 \quad (3.21)$$

where

$$V_{--i} = K_{ij}\Gamma_{--}^j + K_{i\tilde{j}}\tilde{\Gamma}_{--}^j$$
$$V_i^{++} = K_{ji}\Gamma^{++j} + K_{i\tilde{j}}\tilde{\Gamma}^{++j}, \quad (3.22)$$

$$\tilde{V}_{--i} = K_{\tilde{i}\tilde{j}}\tilde{\Gamma}_{--}^j + K_{j\tilde{i}}\Gamma_{--}^j$$
$$\tilde{V}_i^{++} = K_{\tilde{j}\tilde{i}}\tilde{\Gamma}^{++j} + K_{j\tilde{i}}\Gamma^{++j} \quad (3.23)$$

are the Noether supercurrents and $L_{\vartheta_I}, L_{\tilde{\vartheta}_I}$ are the Lie derivatives along the vector superfields $\vartheta_i, \tilde{\vartheta}_i$ respectively. As in the ordinary case [10,11,16] the Poisson-Lie symmetry conditions are satisfied on the extremals, namely

$$(L_{\vartheta_I}K)_{--}^{++} = f_i^{jk}V_j^{++}V_{--k}$$
$$(L_{\tilde{\vartheta}_I}K)_{--}^{++} = \tilde{f}_i^{jk}\tilde{V}_j^{++}\tilde{V}_{--k} \quad (3.24)$$

where $f_i^{jk}$, $\tilde{f}_i^{jk}$ are the structure constants of the Lie superalgabras $g_2$ and $g_1$ respectively. We note that the equations (3.21) are equivalent to the zero supercurvature equations for the $F_{--}^{++}$, $\tilde{F}_{--}^{++}$ components of the super stress tensor $F_{MN}$, namely

$$(F_{--}^{++})_i = D^{++}V_{--i} + \partial_{--}V_i^{++} - f_i^{jk}V_j^{++}V_{--k} = 0$$
$$(\tilde{F}_{--}^{++})_i = D^{++}\tilde{V}_{--i} + \partial_{--}\tilde{V}_i^{++} - \tilde{f}_i^{jk}\tilde{V}_j^{++}\tilde{V}_{--k} = 0$$

(3.25)

These equations are the (4,0) supersymmetric generalization of Poisson-Lie symmetry conditions [10] where the Neother supercurrents (3.22) and (3.23) are the generators of the Lie superalgebra $g_1$ while the structure constants in (3.25) correspond to the Lie superalgebra $g_2$ which is Drinfeld's dual to $g_1$ [17]. Finally, it would be interesting to investigate the connection between quantum groups and the (4,0) supersymmetric WZNW model, which is summarized in the ordinary case by Drinfeld's construction [18]. Furthermore, the geometry and the representation theory of quantum groups, which govern the duality of the quantum D-branes is under study [19].

4 - Conclusion

In this work, we have considered a class of deformed (4,0) supersymmetric principal models where the non-abelian duality of this class connects the SU(2) principal (4,0) supersymmetric model and the O(3) (4,0) supersymmetric sigma model. Furthermore, the non-abelian dual of the O(3) (4,0) supersymmetric sigma model is the extended particular non-abelian dual case of the coset examples discussed in [20]. On the other hand, we have constructed a (4,0) supersymmetric WZNW model by using the analytic superfield G, which takes values in a Lie group G and where its Lie algebra is equipped with ad-invariant non-degenerate inner product. The (4,0) supersymmetric Polyakov-Weigman formula is then obtained by taking into account the kinematics relation. Moreover, we have studied the Poisson-Lie T-duality in the (4,0) supersymmetric WZNW model by considering the correspondence between the complex Manin triple ($g^c$, $g_1$, $g_2$), which is equipped with a hermitian conjugation $\tau$, and the real algebra $g$ with non degenerate invariant inner product and skew-symmetric complex structures.

Acknowledgments

The authors would like to thank Professor K. S. Narain for the interesting discussions and for reading the manuscript and Professor M. Virasoro, the International Atomic Energy Agency and UNESCO for hospitality at the Abdus Salam International Centre for Theoretical Physics, Trieste. This work is supported by the program PARS n$^0$ phys. 27.372/98 CNR and by the framework of the Associate and Federation Schemes of the Abdus Salam International Centre for Theoretical Physics, Trieste, Italy.